\documentstyle[12pt]{article}
\newcommand{\spur}{\mathop{\rm Sp}\nolimits}
\begin{document}

\begin{center}
TRAPPED BOSE--CONDENSATE IN GRAVITY FIELD\\[2mm]
D.B.Baranov, V.S.Yarunin\\[1mm]
{\it Joint Institute for Nuclear Research, Dubna, 141980, Russia}\\[1mm]
{\it yarunin@thsun1.jinr.ru }\\[3mm]
\end{center}

\vspace{1cm}

\begin{quote}
The $1D$ and $2D$ Bose-condensation of trapped atoms in a gravitational
field are considered. The deformation of the finite parabolic potential
in this field is modeling via the combination of two rectangular
$1D$ and $2D$ traps, for which the cut-off
and the re-definition of spectrum are taken into account.
A Bose-condensation $T_c$ shift by the gravity is calculated.
A sign and a magnitude of it in a deformed trap depends on the order of
including the gravitation field.
The special choice of this order may describe three consistent
Bose--condensations with different
temperatures. These transitions may be associated with a transportation
of a trap on the cycle (I) Earth--(II) Space--(III) Earth.


\end{quote}

\vspace{1cm}

The news on the critical temperature $T_c$ of the Bose-Einstein condensation
(BEC) entered the last decade. It is known, that the
critical $T_c$ of the superfluid $^4$He (the oldest -- though inderect --
example of BEC) is the same for any tube with a liquid helium.
The decreasing of $^4$He $T_c$ in porous glasses was found
experimentaly [1] and was explained theoreticaly [2,3], and this shift of a
critical temperature depends on the parameters of a porous glasses [4].
Meanwhile the calculation of BEC $T_c$ in an infinite volume for bosons
with a weak interaction between them showes the increasing of $T_c$ [5,6].

As to the ideal gas, it was found [7], that in an external field,
restricting an infinite volume with this gas, $T_c$ depends on the form of
a field.
The experimental discovery of alkali atoms BEC in magnetic traps [8,9]
gives the opportunity to discuss the same properties in finite systems.
A finite volume  assumes mesoscopic boundary conditions,
which require a new formulation of the thermodynamical limit in theory [10].
The decreasing of an ideal gas BEC $T_c$ was shown [11]
as a finite size correction $N\sim \infty \rightarrow N<\infty$
for the "trap version" thermodynamical limit
$\hbar\omega/T\rightarrow 0$, $N\rightarrow\infty$ calculations.

Here we intend to look for a change of BEC critical temperature $T_c$
via the influence of a gravitational field on $T_c$ of a BEC gas in traps.
A problem of the gravity influence on the
critical temperature of $^4$He arises
in the connection of Space experiments on this subject, and
the same experiments on BEC in trapped atomic gases
are expected in the nearest future [12]. So it is interesting
to predict theoretically the contribution of gravity field
to the  critical BEC temperature in traps.
We deal with the $1D$ and $2D$ theory of spin-less non-interacting atoms
in a finite trap. The basic parameters are taken from the
BEC of alkali atoms experiments [8,9].
We start with the isotropic parabolic trap with a potential barrier $U_0$,
frequency $\omega $ and a height $h$ and introduce two rectangular traps
in order to describe the main part of gravity effect via the deformation
of a trap potential.

The potentials of a parabolic trap are different for the cases
of a zero- and non-zero gravity field in direction $z$.
The potential inside of a trap may be written as
\begin{equation}
U(z) =U_g+(m\omega^2/2)\left(z+\Delta\right)^2,
\quad -h/2<z<h/2, \quad \Delta=g/\omega^2,
\end{equation}
where $m$ is the mass of an atom, $g$ is the free fall acceleration constant.
The dashed line is an initial parabolic potetial,
a solid line is a parabolic potential, deformed by a gravity field.
The gravity field shifts the minimum of potential from
its center $z=0$ by the value  $\Delta$  and down by the value
$U_g=-mg^2/2\omega^2$. The difference between new (shifted) potential
barriers  $U_+$ (on the right) and $U_-$ (on the left)
$$
U_{\pm} = \frac{m\omega^2}{2}\left( \Delta\pm \frac{h}{2}\right)^2 + U_g,
\quad U_0-U_- =U_+ - U_0 =\frac{1}{2}mgh
$$
is equal to $mgh$.
It follows from these formulas that the condition of macroscopic
stability of the trap in gravity field is  $g<\omega^2 h/2$. It is fulfilled
for the experiments  [8,9] and the last inequality
will be satisfied for the frequencies more than 100 $Hz$.
For all that the quantum dynamics  of atoms  undergoes nonperturbate
gravitational disturbance because of the big shift of the trap potential
$\left(U_g\gg \hbar \omega\right)$ in comparison with its frequency.

One-particle boson eigenfunctions $u_n$ and their energies $\varepsilon_n>0$
for the $2D$-case
\begin{equation}
\left\{\nabla^2+\frac{2m}{\hbar^2}\left[\epsilon_n -U(r)\right]\right\}u_n=0,
\quad r=\{x,z\}, \quad u_n(r)=u_{n_x}(x) u_{n_z}(z),
\end{equation}
$$
\epsilon_{n}=\frac{1}{2m}\int \nabla u_n \nabla u_{n}dr+
\int\,U(r) u_n u_{n}dr,\quad n=\{n_x,n_z\}
$$
are the starting points of the quantum analisys of any finite trap $U(r)$,
a ground level of it is $\varepsilon_0\ne 0$.
In order to introduce the Gibbs statistics
we use the path-integral device, applied earlier [13,14] for the BEC
of interacting bosons.

The system of $N$ bosons (2) in a volume $V$ may be represented by the
Hamilton function $H$
$$
\hat H\rightarrow H=\int dr\left[\frac{\hbar^2}{2m}\left(\nabla\psi^*(r,t)
\nabla\psi(r,t)\right)+\psi^* U(r) \psi\right].
$$
Here $\psi^*, \psi$ are the path-integral trajectories with periodical
boundary conditions on $[0,\beta]$. Following [15,16], we
separate the trajectories $b_0^*, b_0$ as the "slow" variables
relative to the "fast" trajectories $b_n^*, b_n,\,n\ne 0$ for $T<T_c$
$$
\psi^*(r,t)=
\frac{1}{\sqrt V}\left(\sum_{n\ne 0}b_n^* (t) u_n (r)+b_0^*(t) u_0(r)\right),
\quad
|b_0|=\sqrt{N_0}\gg 1,
$$
so that the function $N_0(t)$ is the path-integral image of the
"Bose-condensate" particles in Bogoliubov theory [15].
The "broken gauge symmetry" is known as a property
of the bose-gas with the separated condensate fraction.
Still the quasiclassical integral of motion $\bar N=N_0 + \bar {N_1}$
for the system (2) may be expressed by the equation [13]
\begin{equation}
\frac{d\bar N}{dt} \equiv
\{ \hat H, N_0 \} + i[\hat H,\hat {N}_1]=0,\quad
\hat{N}_1=\sum_{n\ne 0}b_n^+ b_n.
\end{equation}
Here $N_1$ is a number of non-condensate bosons,
$\{,\}$ is the Poisson bracket in $b_0, b_0^*$ classical amplitudes
and $[,]$ is the quantum commutator via operators $b_n^+, b_{n'}$,
$n,n'\ne 0$.
The action $S$ with the kinetic term $K$ for $N$ non-interacting bosons
in any volume $V$ looks like
$$
S=\int_{0}^{\beta}(K-H)\, dt
=-\int_{0}^{\beta}\left[ b_0^* \frac{db_0}{dt}+\varepsilon_0 b_0^* b_0
+\sum_{n\ne 0} \left(b_n^*\frac{db_n}{dt}+\varepsilon_0 b_n^* b_n\right)
\right]dt.
$$
An effective action $S_{ef}(\rho,\mu)$ for a condensate
density $\rho=N_0/V$ and a chemical potential $\mu$ is defined via the
partition function with constraint
$$
Q =\int  e^{S_{ef}(\rho, \mu)}\,d\rho\,d\mu=
\spur \left[ \exp(-\beta H) \delta_{R,R_1+\rho}\right],
\quad R=N/V,\quad R_1=N_1/V.
$$
The equation (3) is the background for the definition of $\mu$, and by
the use of Fourier decomposition for $\delta_{R,R_1+\rho}$ with
$\mu=iy/\beta$ we represent $Q$
$$
Q=\int Db_0^*\,Db_0 \prod_{n\not=0} \int D b_n^* D b_{n}
\int_{-\pi}^{\pi} dy \exp [iy (R_1 +\rho - R) + S/\hbar]
$$
as the integral over $y$, the functional integral over the "slow"
$b_0, b_0^*$ and "fast" functional integral trajectories
$b_n, b_{n'}^*$ with the periodical boundary conditions in [$0,\beta$].
The calculation of $Q$ over "fast" variables gives the effective action
$S_{ef}$ of the condensate
$$
Q=\int Db_0^* Db_0\,\int d\mu \exp S_{ef},
$$
$$
S_{ef}=\int_{0}^{\beta}\left[-b_0^*\frac{db_0}{dt}+
\rho (\mu-\varepsilon_0)-\mu R \right]dt -
\sum_{n\ne 0}\ln\left(1-\exp[-\beta(\varepsilon_{n}-\mu)]\right).
$$
It leads to the variational equations $\delta S_{ef}(b_0,b_0^*,\mu)=0$ in
the form
\begin{equation}
\frac{\partial S_{ef}}{\beta\partial\mu}=
\rho+\frac{1}{V}\Sigma-R=0,\quad
\Sigma =\sum_{n_x,n_z\ne 0} f(\epsilon_{n_x,n_z}, \beta, \mu),
\end{equation}
$$
f(\epsilon_{n_x,n_z}, \beta, \mu)\equiv f=
\frac{1}{\exp[\beta(\varepsilon_n-\mu)]-1},
\quad n=\{n_x,n_z\},
$$
\begin{equation}
\frac{\partial S_{ef}}{\beta\partial b_0^*}=
-\frac{db_0}{dt}-b_0 (\varepsilon_0-\mu)=-Lb_0=0,\quad
\mu= \varepsilon_0 - kT \ln\left(1+\frac{1}{N_0}\right).
\end{equation}
Equation (4) means the balance
of condensate and non-condensate particles. The solution of equation (5)
follows via the calculation [13]
$$
(\beta L)^{-1}=[\exp \beta(\epsilon_0 - \mu)-1]
$$
of the operator $L^{-1}$ with the periodical boundary conditions.
The equations (4,5) determine an BEC solution for the model (2,3)
below the critical temperature
for the case of translation-noninvariant systems.
So, the critical temperature $T_c$ for a finite-size trap is determined
by the equations
$\Sigma (T_c)=N$, $\rho=0$, while at the temperarure $T\sim 0$ the equations
$\rho\sim R, R_1\sim 0,\, \mu\sim \varepsilon_0$ are present.
Note, that if we take $b_0^*, b_0$ as $constants$, we get only $T=0$
limit of solutions for $\mu$, it is a property of the ideal Bose-gas.
The formulas (3-5) are valid for any trap $U(r)$.
We apply them for a parabolic trap in formulas (6-8).

The best way to catch a mesoscopic effect is to look for the shift
of a critical temperature $T_c$ of BEC in a trap. It is obvious,
 that a finite size $h$ for a parabolic trap  manifests itself
in the following mesoscopic properties: \\
($i$) a finite number of levels $n_{max}=U_0/\hbar \omega$
of the initial symmetrical parabolic trap,\\
($ii$) a diminishing of the left-side potential barrier
$U_0 \rightarrow (U_- - U_g)<U_0$ by the gravitational field. The
re-definition of one-particle matrix elements
$\epsilon_n$ must be done in both these cases.
It means that we must ($i$) cut-off an upper limit $n<n_{max}$ for barrier
$U_0$ and calculate energies $\epsilon^0_n$, as well as ($ii$)
cut-of an upper limit in (2) for barrier $U_-$ and calculate
energies $\epsilon_n^{\pm}$.
So the further calculation will be done with the formula
\begin{equation}
N=N_0+N_1,\quad N_1=
\sum_{n\ne 0}^{n_{max}}f_{n},
\quad \varepsilon_n=\left\{\begin{array}{cc} \varepsilon^0_n, & g=0, \\
\varepsilon^{\pm}_n, & g\ne 0.\end{array}\right.,
\quad n=\{n_x, n_y\}.
\end{equation}
of the full number $N$ of atoms in a trapped ideal bose gas.
The level of $n_{max}$ defines the upper bound states of atoms.
The numerical parameters of the experiments [8,9] for a parabolic trap
in an isotropic version may be reduced
approximately to the values
\begin{equation}
U_0\sim 10^{-9}eV, \quad h=2mm,\quad U_+ -U_- =mgh,
\quad \omega\sim 10^{-13}eV.
\end{equation}
We suppose the critical $3D$ peak density [8,9] $R\sim 10^{14} cm^{-3}$
to be independent on time as the evaporation cooling is neglected
here. The $nD$ scaling for the number of atoms in a parabolic trap
with a volume $V$,
over wich the system is confined, looks like [10]
\begin{equation}
R|_{nD}=\frac{N}{V}, \quad V\sim \left(\frac{m\omega^2}{T_c}\right)^{-n/2},
\quad n=1,2,3.
\end{equation}

In order to symplify a problem we would like to represent
the parabolic trap by the rectangular trap.
It means, that we describe
the deformation of the parabolic potential in the gravitation field
by transforming the symmetrical rectangular trap
with potential $U_0$ to the asymmetrical one with barriers $U_+ - U_-=mgh$.
Our purpose is to estimate
the shift of a critical temperature in the asymmetrical rectangular
trap relative the symmetrical one as a measure of the gravity field
influence on BEC, tunneling effects are not considered.
The  values of $R$ in a $1D$ and $2D$ parabolic traps are $10^5 cm^{-1}$ and
$10^{10} cm^{-2}$ in the view of (8).
We keep these $R$ the same for a rectangular traps.
So a number of atoms in a $1D$ rectangular trap is
$N_1|_{1D}\sim 10^{5}cm^{-1}\times 0.2 cm\sim 2\cdot 10^4$, and in a
$2D$ rectangular trap is $N_1|_{2D}\sim 10^{10}cm^{-2}\times (0.2 cm)^2\sim
4\cdot 10^8$. The equation for a spectrum $\varepsilon_n\rightarrow E_k$
for the symmetrical $1D$ and $2D$ traps looks like [17]
$$
tg(kh)=\frac{2k\hbar}{p_k^0}
\left[\frac{k^2 \hbar^2}{(p^0_k)^2}-1\right]^{-1},
\quad E_k=\frac{\hbar^2 (k^2_x + k^2_z)}{2m},
\quad p^0_k=\sqrt{2m(U_0-E_k)},\quad k=\{k_x, k_z\}
$$
and for the asymetrical (deformed) trap for $k=k_z$ looks like
$$
tg(kh)=k\hbar \left(\frac{1}{p^+_k}+\frac{1}{p_k^-}\right)
\left(\frac{k^2 \hbar^2}{p_k^+ p_k^-}-1\right)^{-1},\quad
p_k^{\pm}=\sqrt{2m(U_{\pm}-E_k)}.
$$
The distances between the energy levels $E_k$ of an initial rectangular trap
with the given parameters (7) are $\sim 10^{-19}eV$
at a bottom and $\sim 10^{-13}eV$ at a top barrier $U_0$.
It has $\sim 10^5$ levels (contrary $\sim 10^4$
in experiments [8,9]), so the average energy distance
between levels $\sim 7\cdot 10^{-14}eV$ is the same as the frequency
$\hbar \omega\sim 6\cdot 10^{-14}eV$ of the paprabolic trap.

In order to get more systematic picture we start with 1D case with the
$R\sim 10^5 cm^{-1}$ density.
The critical temperature $T_c^0|_{1D}=2.41\cdot 10^{-10} K$ for an initial
rectangular trap $U_0$ is found from the equation (6) with the use
of calculated spectrum $E_k$
$$
N_0=0,\quad N_1^0|_{1D}=N|_{1D}=
\sum\limits_{k_z\ne 0}f(T_c^{0}|_{1D}, E_{k_z}^{0})= 2\cdot 10^4,
\quad k<k_{max}.
$$
A suitable accuracy of calculations may be received
for $k_{max}\sim 1000$ in the last sum in $k_z$ variable.
The deformed rectangular trap $U_-\sim 10^{-8}U_0$ is determined via
the parameters $h$ and $U_+ - U_- =mgh$ in (7).
It contains only $\sim 10$ levels, which are occupied by
$$
N_1^{\pm}|_{1D} =
\sum\limits_{k_z\ne 0}^{10} f(T_c^0|_{1D}, E_{k_z}^0)=1.8\cdot 10^4 <N|_{1D}
$$
atoms. These atoms undergoes a new BEC transition in the system of
re-defined levels with a critical temperature in a deformed trap
$T_c^{\pm}|_{1D}=2.27\cdot 10^{-10}$, that is found from the equation
$$
N_0=0,\quad N_1^{\pm}|_{1D}=
\sum\limits_{k_z\ne 0}^{10} f(T_c^{\pm}|_{1D}, E_{k_z}^{\pm})= 1.8\cdot 10^4,
\quad k<k_{max}.
$$
So the shift of the critical temperature is $\Delta(T_c)|_{1D}=
T_c^{\pm}|_{1D}-T_c^0|_{1D}=-0.14\cdot 10^{-10}$. The rest part
$0.2\cdot 10^4 $ of atoms becomes non--trapped (continuous
spectrum).

As to the $2D$ case, the sum over the variables ${k_x, k_z}$
is taken within the constraint condition \footnote{This condition is
trivial for an initial trap, but rather effective for a deformed trap.}
$E_k< E(n|_{max})$, $k=\{k_x, k_z\}$.
A temperature  $T^0|_{2D} =5.05\cdot 10^{-7} K$ for an initial
rectangular trap $U_0$ is found from the equation (6)
$$
N_0=0,\quad N_1|_{2D}=N|_{2D}=
\sum\limits_{k_x,k_z\ne 0}f(T_c^0|_{2D}, E_{k_x k_z}^{0})= 4\cdot 10^8,
\quad k<k_{max}.
$$
The latter sum includes all the items of the partition function within
the possible degeneracy of levels.
A suitable accuracy of calculations may be received
for $k_{max}\sim 400$ in the last sum in $k_x, k_z$ variables.
The deformed rectangular trap $U_-\sim 10^{-8}U_0$
contains $\sim 10$ levels, which are occupied by
$$
N_1^{\pm}|_{2D}=
\sum\limits_{k_x,k_z\ne 0}^{10}f(T_c^0|_{2D}, E_{k_x k_z}^{0})=
0.96\cdot 10^8<N|_{2D}
$$
atoms. Just like in $1D$ case, we look for a transition temperature
in a deformed trap using an equation
$$
N_0=0,\quad N_1^{\pm}|_{2D}=
\sum\limits_{k_x,k_z\ne 0}^{10}f(T_c^{\pm}|_{2D}, E_{k_x k_z}^{0}).
$$
The result of calculation is $T_c^{\pm}|_{2D}=5.03\cdot 10^{-7} K$,
so the shift of a critical temperature is
\begin{equation}
\Delta(T_c)|_{2D}=T_c^{\pm}|_{2D}-T_c^0|_{2D}=-0.02\cdot 10^{-7}K.
\end{equation}

Now we describe the $T_c$ shift in a system with the same number
of atoms, $4\cdot 10^{8}$ for example (for $2D$-case in our theory).
The critical temperature
for the symmetrical trap was found above as $T_c^0|_{2D}= 5.05\cdot 10^{-7} K$.
The critical temperature for the asymmetrical trap
$\left( T_c^{\pm}|_{2D}\right)^* = 2.02 \cdot 10^{-6} K$
is found from the equation
$$
N_0=0,\quad N_1|_{2D}=N|_{2D}=
\sum\limits_{k_x,k_z\ne 0}^{10}
f\left[ \left( T_c^{\pm}|_{2D}\right)^*, E_{k_x k_z}^{\pm}\right]
=4\cdot 10^8,
$$
\begin{equation}
\Delta(T_c)= \left( T^{\pm}_c|_{2D}\right)^*
-T_c^0|_{2D}=1.52\cdot 10^{-6} K.
\end{equation}
Note, that  $\left( T_c^{\pm}|_{2D}\right)^* $ in (10) does not coinside
with the $T_c^{\pm}|_{2D}$ in (9), the reason is that the processes
(9) and (10) are different. This important fact will be discussed in the
end of the paper.

The above calculations concern a rectangular trap.
As to an initial parabolic trap, it has the same barriers $U_{\pm}$ with
the same shift $U_+ - U_-=mgh$, but differs in spectrum.
It is reasonable
to suppose, that the same scale of a BEC critical temperature shift will
be valid for the parabolic trap in the gravitation field,
so the qualitative correspondence between the critical temperatures
$T_c^0\to T|_{g=0}$, $T_c^{\pm}\to T|_{q\ne 0}$ for a rectangular and
parabolic traps is present. Meanwhile the tunneling effect
is not reproduced by a rectangular trap.
As it is seen from the formula (1), the macroscopic stability of a
parabolic trap
in a gravity field is expressed by the unequality $\omega^2>2g/h$.
It is clear, that this condition is trivial in the absence of gravity
field or for the trap on the whole axe $-\infty<z<\infty$.
The finite size trap effect in
the case of a deformed rectangular trap $n_{max}\sim 10$, attributed
to the most noticable gravity shift of $T_c$,
may be reproducrd for the parabolic trap via the condition
$$
(\omega^2 h-2g)\rightarrow +0.
$$
This condition means, that the largest shift of $T_c$ may be noticed
just before the destruction of the parabolic trap.

Let us turn in a conclusion to the non-ideal gas
with an interaction $G(r,r')$ beteen atoms.
In the 1D case the effective action in (4,5) for interacting
bosons is generalized in the Hartree-Fock-Bogoliubov approximation
as [13]
$$
S_{ef}(\rho,\mu)=-\beta\rho^2 \gamma_0 V+\beta\mu(\rho - R)V
-\beta\rho \varepsilon_{0}+
$$
$$
\frac{1}{2}\sum_{nn'\ne 0}\left[\beta (W_{nn'}-  \mu\delta_{nn'})-
4 ln\, sh \frac{\beta E_{nn'}}{4} + \beta A_{nn'}\right],\quad
A_{nn'}=-\frac{V\phi_n^* M_{nn'}m \phi_{n'}}{E^2_{nn'}},
$$
$$
\rho=\frac{|b_0|^2}{V},\quad
m=\left(\begin{array}{cc}
-1&0\\0&1\end{array}\right),\quad
M=\left(\begin{array}{cc}W_{nn'}-\mu\delta_{nn'} & 2\gamma_{nn'} \rho\\
-2\gamma_{nn'} \rho & -W_{nn'}+\mu\delta_{nn'}\end{array} \right),
$$
$$
W_{nn'}=w_{nn}/V+2\rho\gamma_{nn'},\quad
E_{nn'}^2=(M|_{11})^2-(M|_{12})^2, \quad
\phi_{n}^*=(b_0^*, b_0)(\rho\gamma_{0n}).
$$
Matrix elements of  $G(z,z')$
in $\delta$--approximation  $G(z,z')\rightarrow G\cdot \delta (z-z')$
$$
\gamma_0=\frac{G}{2} \int dz u_0^4,\quad \gamma_{0n}=
\frac{G}{2}\int dz u_0^3 u_n,\quad
\gamma_{nn'}=\frac{G}{2}\int dz u_0^2 u_n u_{n'}
$$
are detrmined for the functions of harmonic oscillator in $z$ direction
with the potential barrier  $(U_- - U_g)<U_0$ at the left endpoint
of the trap. Only the cut-off spectrum contribution
(without the re-definition of $\epsilon_n$) is taken into account.
Equations $\delta S_{ef}(\rho,\mu)=0$  for a trap with  the constant
density  of atoms
$R=N/V$  may be evaluated approximately by dividing the sum over
full number of trap levels into two terms with
 $n\ll n_0$ and $n\gg n_0$. Parameter $n_0\simeq 100$
is defined under the condition
\begin{equation}
n_0 \hbar \omega \simeq \rho \gamma_{n_0 n_0}\simeq GN_0,
\quad N_0\sim N,
\end{equation}
following from the unequality
$$
10^{-13}eV \simeq \hbar \omega \ll GN_0\simeq 10^{-11}eV,\quad
G=  \frac{4\pi \hbar^2 a}{m},
$$
where the energy of interaction between atoms is taken from the
estimation [8] for the scattering length   $a=4.9 nm $
with density of atoms $R\sim 10^{10} {cm}^{-3}$.
Thus, a density of condensate is written in the form
$\rho=\rho^< + \rho^> $, where
$\rho^<=\rho|_{n\ll n_0}$, $\rho^>=\rho|_{n\gg n_0}$.
We can estimate the ratio $Y$ of Bose-condensate densities in a
trap with  $\rho|_g$  and without gravity $\rho|_0$
$$
Y=\frac{\rho|_g}{\rho|_0}=\frac{(\rho^< +\rho^>)|_g}{(\rho^< +\rho^>)|_0}
\simeq \frac{\rho^<|_g}{\rho^<|_0}
\left(1-\frac{\rho^>|_0}{\rho^<|_0} + \frac{\rho^>|_g}{\rho^<|_g}\right),
\quad \rho^<|_{0,g}\gg \rho^>|_{0,g}.
$$
In the case of a "strong" gravity field $(U_- -U_g)\ll U_0$ the inequality
$\rho^<|_g > \rho^<|_0$ follows after the evaluations.
In the case of a "weak" gravity field
$(U_- - U_g)\sim U_0$ the equations $\rho^<|_g=\rho^<|_0$ and
$\rho^>|_g > \rho^>|_0$ are valid. The inequalities
$$
Y_{strong}=
\frac{\rho^<|g}{\rho^<|_0}\left(1-\frac{\rho^>|_0}{\rho^<|_0}\right)>1,
\quad Y_{weak}=\left(1-\frac{\rho^>|_0 - \rho^>|_g}{\rho^<|_0}\right)>1
$$
are proved. Taking into account, that the large Bose-condensate density
corresponds the large critical temperature, we note,
that the gravitional field increases the critical temperature
$T_c$ of Bose-condensation for non-ideal gas in a trap.
The parameter $n_0$ in (11) corresponds the parameter
$n_{max}$ in (6) as the border of BEC occupated states.
The tunneling effects are also neglected here.

We would like to look for a correspondence  between formulas (9,10) and
 the probable motion of  a trap between the Earth and Space. We measure the
$\left( T_{c}^{\pm}|_{2D}\right)^* =2.02\cdot 10^{-6} K$
temperature for the $N|_{2D}$ atoms on the Earth
in a deformed trap in the begining. Then the trap goes to Space,
and the same number of atoms $N|_{2D}$ undergoes the phase transition
(in a non--deformed trap) at the temperature
$T_c^{0}|_{2D}=5.05\cdot 10^{-7} K$ with
a shift (10). If the trap goes back to the Earth, some part of atoms
becomes non--trapped due to its deformation. The rest part $N_1^{\pm}|_{2D}$
of them undergoes a phase transition at the temperature
$T_c^{\pm}|_{2D}=5.03\cdot 10^{-7} K$,  and a temperature shift
is given by (9).

So we have three transition temperatures (with the same trap!)
in different cases I, II, III
\begin{equation}
\left( T_c^{\pm}|_{2D}\right)^*
=2.02\cdot 10^{-6} K\to T_c^{0}|_{2D}=5.05\cdot 10^{-7} K
 \to  T_c^{\pm}|_{2D}=5.03\cdot 10^{-7} K,
\end{equation}
\begin{center}
start on the Earth (I) $\to$ in Space(II) $\to$ finish on the Earth (III).
\end{center}
This is a picture for an ideal Bose--gas. The sign of $T_c$ shift
for the non--ideal gas on the stage I$\to$ II was performed in the last
formulas of the paper, but the re--definition of levels was not taken
into account. It seems, that the $3D$ calculation will increase
the magnitudes of the critical temperatures in (12).

The mesoscopic nature of the problem considered above is seen in
the strong dependence of the calculated $T_c$ shifts on the
initial parameters, and the way of their association with  experimental
prototypes.\\

[1] G.K.S.Wong, P.A.Crowell, H.A.Cho and J.D.Reppy, Phys. Rev. Lett.
{\bf 65} 2410,

(1990).

[2] K.~Huang, H.-F.~Meng, Phys. Rev. Lett. {\bf 69}, 644 (1992).

[3]  V.~Sa-yakanit, V.~Yarunin, P.~Nisameneephong,
Phys. Lett. {\bf A 237}, 152 (1998).

[4] Int. Symp. on Ultra-Low Temp. Phys., Aug.12-15, 1999,
St.Petersbourg, Russia.

[5] K.~Huang, cond-mat/9904027 1 Apr 1999.

[6] G.~Baym, J.-P.~Blaziot, J.~Zinn-Justin, cond-mat/9907241 16 Jul 1999.

[7] V.~Bagnato, D.~Pritchard, D.~Kleppner, Phys. Rev. A {\bf 35},
4354 (1987).

[8] K.B.~Davis, M.O.~Mewes, M.R.~Andrews, N.J.van ~Druten, D.S.~Durfee,
D.M.~Kurn

and W.~Ketterle, Phys. Rev. Lett. {\bf 75} 3969 (1995).

[9] M.O.~Mewes, M.R.~Andrews, N.J.van Druten, D.M.~Kurn, D.S.~Durfee

and W.~Ketterle, Phys. Rev. Lett. {\bf 77} 416 (1996).

[10] K.~Damle, T.~Senthil, S.~Majumdar, S.~Sachdev,
Evrophys. Lett., {\bf 36}, 7 (1996).

[11] L.~Pitaevskii, Usp. Fiz. Nauk, {\bf 168} 641 (1998);

cond.mat/9806038 v 2 12 Okt 1998.

[12] Roadmap on "Fundamental Physics in Space", Preprint NASA, JPL
400-808,

4/99.

[13] V.~Yarunin, Teor. Mat. Fiz. {\bf 96} 37 (1993), {\bf 109}
295 (1996); V.~Yarunin and L.~Siurakshina, Physica {\bf A 215} 261 (1995);
V.~Yarunin, Teor. Mat. Fiz. {\bf 119} 308 (1999);
V.~Yarunin, Fiz. Nizk. Temp. {\bf 24} 176 (1998).

[14] V.~Yarunin, D.~Baranov. Preprint JINR E17-99-172 (To be publ.
in J.Low Temp. Phys., USA, 2000).

[15] N.N.~Bogoliubov, Isvestia AN SSSR, Ser. Fiz., {\bf 11} 77 (1947).

[16] V.N.~Popov, L.D~Faddeev, J. Exper.i Teor. Fiz., {\bf 4} 1315 (1964).

[17] S.~Flugge, Practical Quantum Mechanics, v.1, Springer-Verlag, 1971.\\

\end{document}